\documentclass[prb,fleqn,12pt,showpacs]{revtex4}
\usepackage{epsfig}
\usepackage{graphicx}
\usepackage{amsfonts}
\begin{document}
\title{Magnetic excitations in iron pnictides}
\author{Nimisha Raghuvanshi, Sayandip Ghosh, Rajyavardhan Ray, Dheeraj Kumar Singh, and Avinash Singh}
\email{avinas@iitk.ac.in}
\affiliation{Department of Physics, Indian Institute of Technology Kanpur - 208016}
\begin{abstract}
Spin wave dispersion and damping are investigated in the metallic SDW state of different itinerant electron models including a small interlayer hopping. Magnetic excitations in iron pnictides are shown to be well understood in terms of physical mechanisms characteristic of metallic magnets, such as carrier-induced ferromagnetic spin couplings, intra-band particle-hole excitations, and the spin-charge coupling mechanism, which is also important in ferromagnetic manganites. 
\end{abstract}
\pacs{75.30.Ds,71.27.+a,75.10.Lp,71.10.Fd}
\maketitle
\newpage

\section{Introduction}
Single-crystal neutron scattering studies of iron pnictides have indicated a commensurate magnetic ordering of iron moments ordered ferromagnetically in the $b$ direction and antiferromagnetically in the $a$ and $c$ directions.\cite{goldman_2008} Inelastic neutron scattering measurements\cite{zhao_2008,diallo_2009,zhao_2009} in $\rm A Fe_2 As_2$ (A = Ca, Ba, Sr) yield well-defined spin-wave excitations up to the zone boundary on an energy scale $\sim 200$meV. The realization of a $(\pi,0,\pi)$ ordered SDW state has opened the possibility of observing phenomena in this class of compounds which are characteristically associated with both the antiferromagnetic (AF) state such as quantum spin fluctuations, hole/electron motion in AF background, spin-fluctuation mediated pairing, as well as the metallic ferromagnetic (F) state such as carrier-induced spin interactions, correlation-induced spin-charge coupling, and non-quasiparticle states. 

The metallic SDW state in principle also allows for spin wave damping due to decay into low-energy intra-band particle-hole excitations, in addition to the usual gapped inter-band (Stoner) excitations. Indeed, from the observed high energy behaviour of spin wave damping ascribed to particle-hole excitations,\cite{diallo_2009} it has been inferred that the full excitation spectrum can not be understood in terms of the localized spin models.\cite{yao_2008,si_2008,applegate_2010,manousakis_2010} Weakly damped spin waves near the ordering wavevector have been obtained within multiband models from the imaginary part of the spin fluctuation propagator.\cite{brydon_2009,cvetkovic_2009,knolle_2010} Spin wave energy renormalization and spectral broadening have also been investigated in the metallic AF state.\cite{met_af} 

Spin wave dispersion in the $(0,\pi)$ and $(0,\pi,\pi)$ ordered SDW states and the role of doping-induced ferromagnetic spin couplings mediated by the exchange of the particle-hole propagator has been investigated within single-band and two-band\cite{raghu_2008} models with the Hubbard interaction\cite{faf0,faf1,faf2} and with a local spin-fermion exchange coupling\cite{lv_2010} as in the ferromagnetic Kondo lattice model for ferromagnetic manganites.

The presence of ferromagnetic chains in the SDW state of iron pnictides also suggests the possibility of spin wave damping due to the spin-charge coupling mechanism as in metallic ferromagnets,\cite{quantum,spch3} which provides a quantitative understanding of the measured spin wave linewidth in neutron scattering studies of ferromagnetic manganites.\cite{zhang_2007,qfklm,qfklm2} The spin-charge coupling mechanism is especially important in saturated ferromagnets where the large inter-band gap forbids any low-energy spin wave damping due to decay into particle-hole excitations. Inelastic neutron scattering studies of iron pnictides indicate a constant ratio $\Gamma_{\bf q}/\omega_{\bf q} \sim 0.2$ of the spin wave linewidth to energy, and clearly do not show any steep increase in the spin wave damping at higher energies as expected from decay into Stoner excitations.\cite{zhao_2009}

In this paper, we will therefore investigate these two mechanisms for spin wave damping in the $(0,\pi)$ ordered SDW state of itinerant electron models. We will also extend our earlier spin wave analysis in the $(0,\pi,\pi)$ ordered SDW state\cite{faf0,faf1} to include hopping anisotropy as appropriate for a layered system, and compare with the measured spin wave dispersion for iron pnictides. For this purpose, we will consider the single-band Hubbard model on a simple cubic lattice, with first and second neighbour hoppings $t$ and $t'$ in the x-y plane and a small interlayer hopping $t_z$. 

We will also study the minimal two-band model with reference to the $\rm d_{xz}$ and $\rm d_{yz}$ orbitals of interest for iron pnictides,\cite{raghu_2008} with the effective hopping parameters $t_1$ - $t_4$ resulting from the hybridization of Fe 3d orbitals with themselves as well as through the As 3p orbitals lying above and below the square plaquettes formed by the Fe atoms. It is convenient to include the inter-orbital Hund's coupling term on an equal footing with the intra-orbital Hubbard interaction term within the general multi-orbital correlated electron model:
\begin{equation}
H = -\sum_{\langle ij\rangle \mu\nu\sigma} t_{ij}^{\mu\nu} 
(a_{i\mu\sigma}^\dagger a_{j\nu\sigma} + a_{j\nu\sigma}^\dagger a_{i\mu\sigma}) 
- \sum_{i \mu \nu} U_{\mu \nu} {\bf S}_{i \mu} \cdot {\bf S}_{i \nu}
\end{equation}
where the interaction matrix elements $U_{\mu \nu} = U_{\mu}$ for $\mu = \nu$ and $U_{\mu \nu} = 2J$ for $\mu \ne \nu$ refer to the intra-orbital and inter-orbital Coulomb interaction terms, respectively. For the two-band model, we will specifically consider the two cases: \\
(i) $t_1$=-1, $t_2$=1, $t_3$=$t_4$=-0.3 at finite hole doping and \\
(ii) $t_1$=-1, $t_2$=1.3, $t_3$=$t_4$=-0.85 at finite electron doping \\
both of which yield stable metallic SDW states on including a finite Hund's coupling.\cite{faf2} Case (ii) yields circular electron and hole pockets near the bottom of the upper band.\cite{raghu_2008}

\section{Spin wave dispersion}

\begin{figure}
\vspace*{-10mm}
\hspace*{0mm}
\psfig{figure=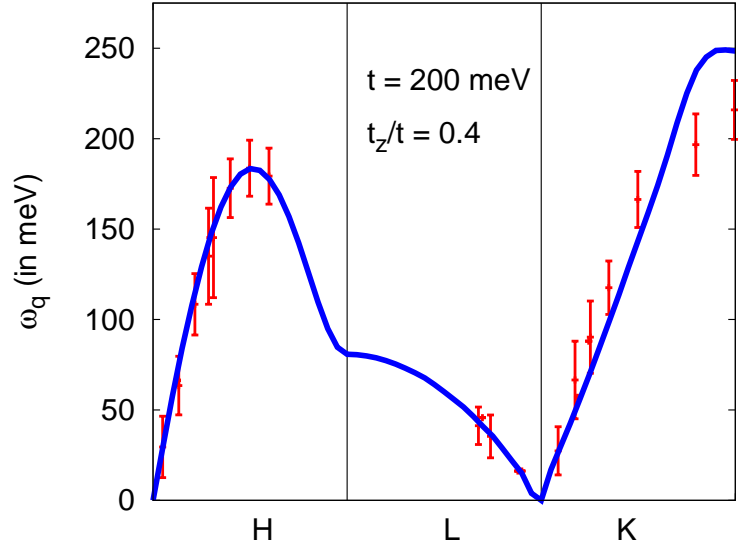,width=75mm,angle=-90}
\vspace*{5mm}
\caption{The calculated spin wave dispersion in the $(\pi,0,\pi)$ ordered metallic SDW state of the Hubbard model with a small inter-layer hopping term and comparison with the neutron scattering data (from Ref. [4]).}
\end{figure}

Figure 1 shows the calculated spin wave dispersion $\omega_{\bf q}$ for the Hubbard model with anisotropic hopping, along with the neutron scattering experiment data points.\cite{zhao_2009} Here H,K,L correspond to wavevector ${\bf q}$ along the x,y,z directions. The overall energy scale is fitted with $t$=200 meV, and the interlayer-to-planar hopping ratio is taken as $r$=$t_z/t$=0.4. The other Hubbard model parameters are $U/t \approx 11$ (corresponding to $\Delta/t=3$ where $2\Delta \equiv mU$), $t'/t=0.3$, and hole doping $x \approx 40\%$, as earlier.\cite{faf1} 

Analytical expressions for spin wave dispersion in the $(0,\pi)$ and $(0,\pi,\pi)$ ordered SDW states were derived earlier in two and three dimensions in the strong coupling limit of the $t$-$t'$ Hubbard model at half filling.\cite{faf0}  It is useful to extend these expressions to approximately include the ferromagnetic coupling induced at finite doping, the finite sublattice magnetization $m$ in the doped state, and anisotropic hoppings in different directions as in iron pnictides. For the $(0,\pi)$ state we thus obtain:
\begin{equation}
\left ( \frac{m \omega_{\bf q}}{J} \right )^2 = \left \{ \left ( 1+\frac{2J'}{J} \right ) - b(1-\cos q_x) \right \}^2 
- \left \{ \left ( 1+\frac{2J'}{J} \cos q_x \right ) \cos q_y \right \}^2
\end{equation}
where $J$ and $J'$ represent first and second neighbour AF spin couplings in the metallic SDW state, which reduce to $4t^2/U$ and $4t'$$^2/U$, respectively, in the insulating limit.\cite{faf0} The coefficient $b < 0$ represents the negative contribution of the carrier-induced ferromagnetic spin couplings in the $x$ direction, whereas $b=1$ in the insulating limit. Similarly, for the three-dimensional $(\pi,0,\pi)$ SDW state of a layered system, we obtain:
\begin{equation}
\left ( \frac{m \omega_{\bf q}}{J} \right )^2 = \left \{ \left ( 1+\frac{2J'}{J} \right ) - \frac{b}{2}(1-\cos q_y) \right \}^2 
- \left \{ \left ( 1+\frac{2J'}{J} \cos q_y \right ) \left ( \frac{\cos q_x + r^2 \cos q_z}{1+r^2} \right ) \right \}^2
\end{equation}
where $r=t_z / t$ is a small interlayer-to-planar hopping ratio. The energy scale on the left should be $J(1+r^2)$ in general. The resulting spin-wave dispersions evaluated from Eqs. (2) and (3) are shown in Fig. 2 with $2J'/J = 2.0$ and $b=-0.3$ for the two-dimensional case, and $2J'/J = 1.0$, $b=-0.3$, and $r^2=0.1$ for the layered system. The dispersion in the latter case provides an excellent description of the neutron scattering results for iron pnictides.

As seen from Fig. 2, the spin wave dispersions calculated from the approximate expressions given in Eqs. (2,3) are remarkably close to that obtained from the $t$-$t'$ Hubbard model, indicating that the essential features of the carrier-induced spin couplings are appropriately incorporated. The substantial enhancement of $J'/J$ as compared to $(t'/t)^2$ reflects the additional second-neighbour AF spin couplings generated at finite doping. 

For the general spin wave propagator structure in the two-sublattice basis, including both retarded and advanced poles, we correspondingly obtain:
\begin{equation}
\chi^{-+}({\bf q},\omega) = -\frac{1}{2}\left ( \frac{J}{\omega_{\bf q}} \right ) \left [ \begin{array}{lr} 
\alpha_{\bf q} - m (\omega/J)  & -\beta_{\bf q}  \\
-\beta_{\bf q} & \alpha_{\bf q} + m (\omega/J) \end{array} \right ]
\left ( \frac{1}{\omega - \omega_{\bf q} + i \eta} - \frac{1}{\omega + \omega_{\bf q} - i \eta} \right )
\end{equation}
where the momentum-dependent terms $\alpha_{\bf q}$ and $\beta_{\bf q}$ represent the terms in the two curly brackets in Eqs. (2) and (3). Including the magnetization $m$ as above ensures that the spin commutation property $\langle [S^+ , S^-] \rangle = \langle 2S^z \rangle $ relating local spin correlations and magnetization is identically satisfied.

\begin{figure}
\vspace*{-0mm}
\hspace*{0mm}
\psfig{figure=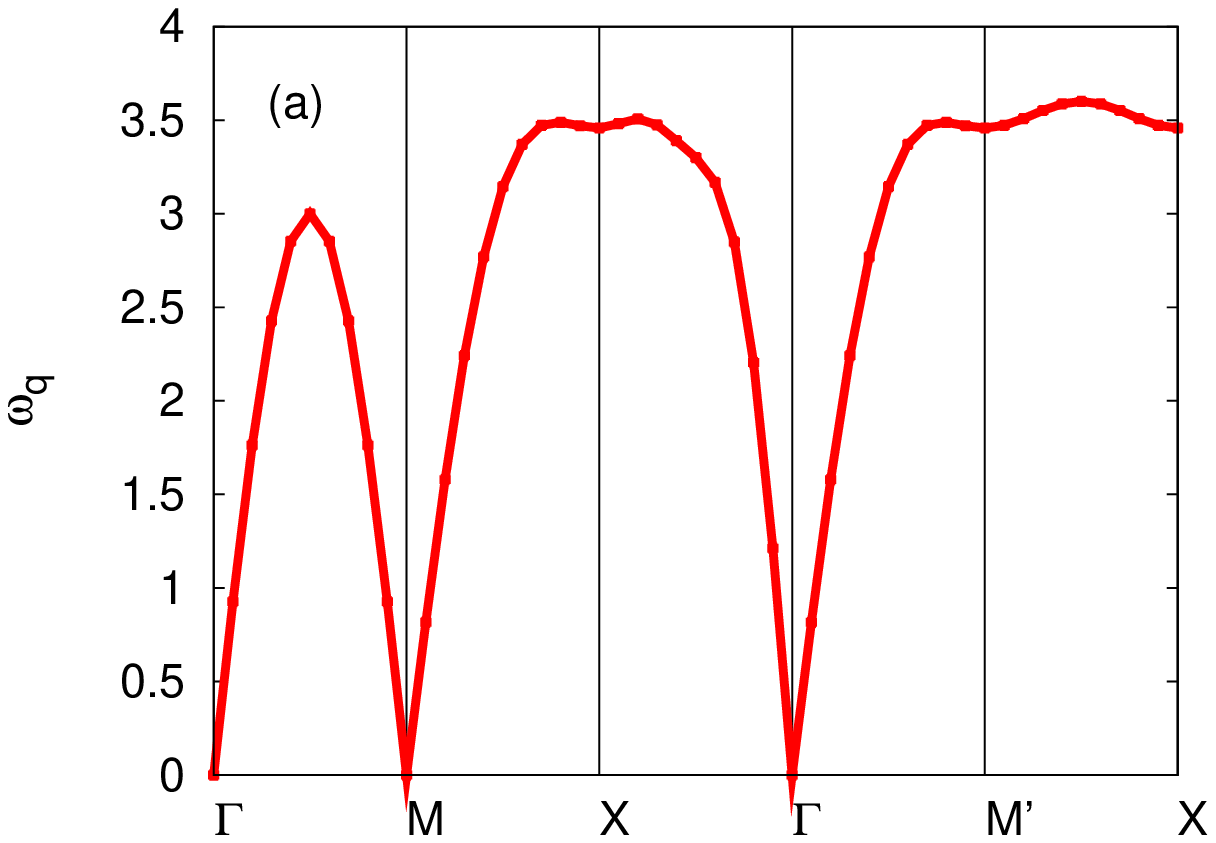,width=75mm,angle=0}
\psfig{figure=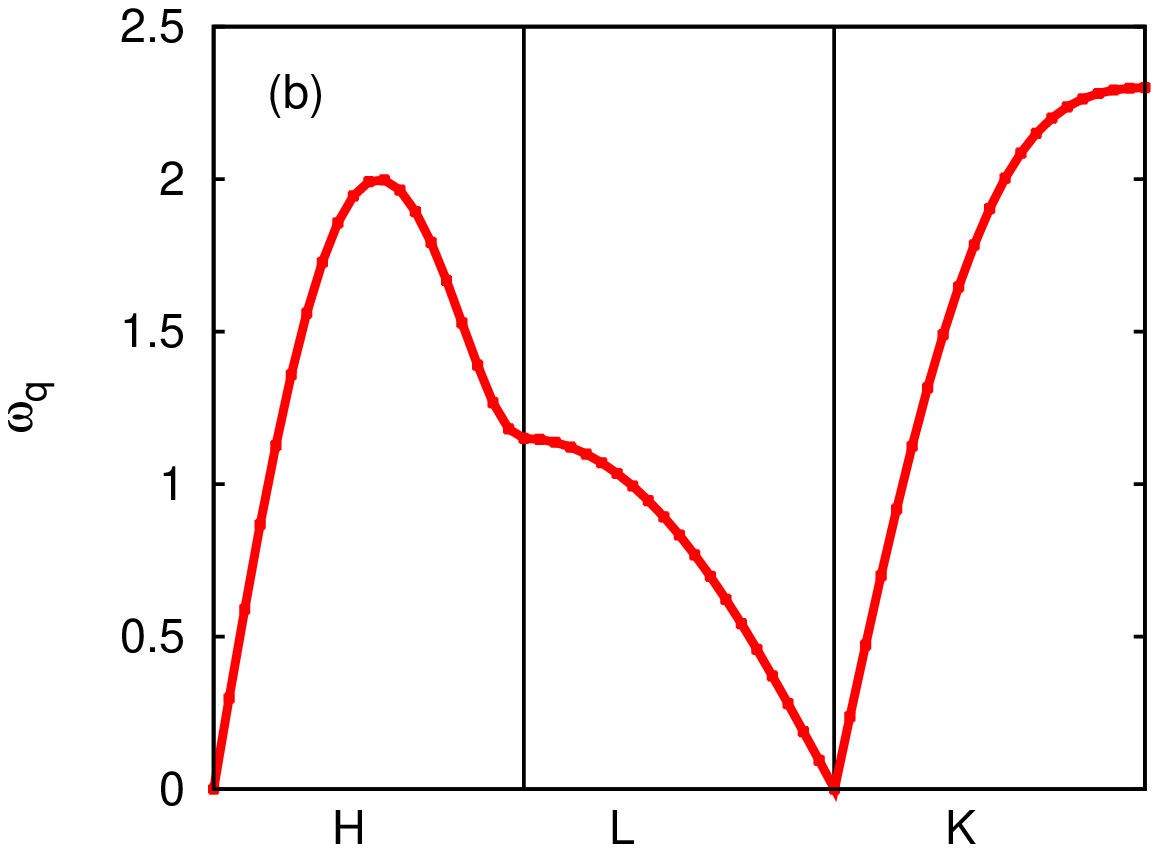,width=75mm,angle=0}
\vspace*{-0mm}
\caption{Spin wave dispersions $\omega_{\bf q}$ (in units of $J/m$) evaluated from the analytical expressions given in Eqs. (2) and (3) for the (a) $(0,\pi)$ and (b) $(\pi,0,\pi)$ ordered SDW states, obtained by including a doping-induced ferromagnetic spin coupling term and a small interlayer hopping term.}
\end{figure}

\section{Spin-wave damping due to decay into intra-band particle-hole excitations}

As mentioned earlier, spin-wave decay into low-energy intra-band particle-hole excitations in principle provide a mechansim for spin wave damping in a metallic SDW state. However, whether the simultaneous requirements of momentum and energy conservation afford sufficient phase space to yield significant linewidth is a relevant question. In the following, we will investigate this aspect from the imaginary part of $\chi^0({\bf q},\omega)$ for the $t$-$t'$ Hubbard model, as well as for the minimal two-band models, both hole doped and electron doped. 

\begin{figure}
\vspace*{-0mm}
\hspace*{0mm}
\psfig{figure=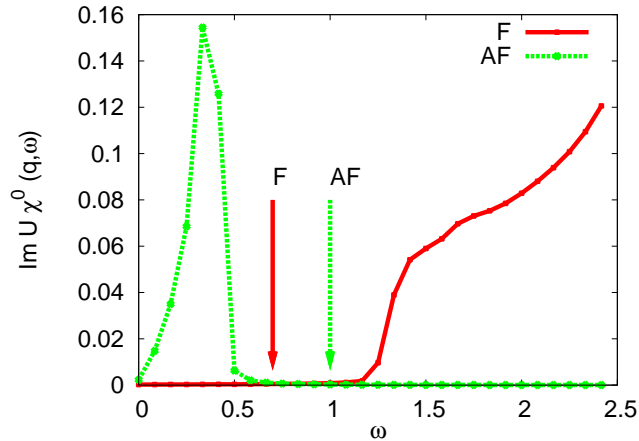,width=85mm,angle=0}
\vspace*{-0mm}
\caption{The frequency dependence of the imaginary part of the particle-hole propagator $\chi^0 ({\bf q},\omega)$, including both intra-band and inter-band particle-hole excitations. The imaginary part vanishes over a large frequency range which includes the spin wave energy for mode 
$\bf q$, implying vanishing contribution to spin wave damping due to decay into particle-hole excitations.}
\end{figure}

Fig. 3 shows the frequency dependence of the imaginary part of the particle-hole propagator $\chi^0 ({\bf q},\omega)$ for the $t$-$t'$ Hubbard model, including both intra-band and inter-band particle-hole excitations. In the two typical cases shown here, with wavevector $q=\pi/2$ in the F and AF directions, the imaginary part vanishes over a large frequency range, which includes the spin wave energies for both modes (shown by arrows), implying vanishing contribution to spin wave damping due to decay into particle-hole excitations. Similar behaviour is obtained for the two-band model with $t_3=t_4=-0.3$, again indicating no intra-band contribution to spin wave linewidth.

\begin{figure}
\vspace*{-0mm}
\hspace*{0mm}
\psfig{figure=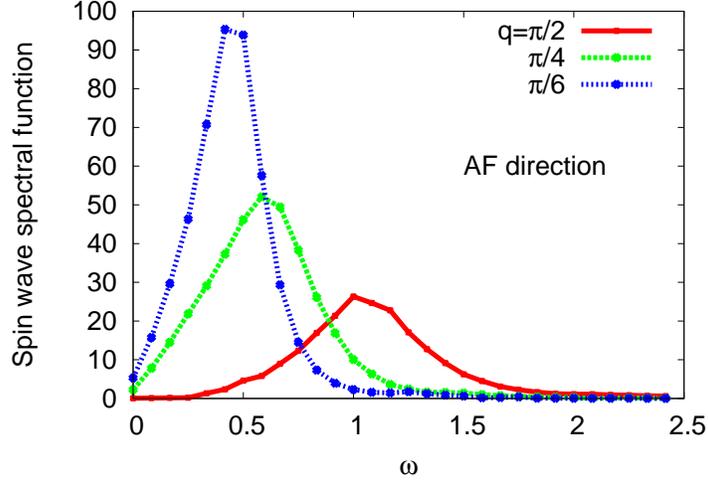,width=95mm,angle=0}
\vspace*{-0mm}
\caption{The spin-wave spectral function for the two-band model with circular electron and hole pockets,\cite{raghu_2008} showing significant broadening for wavevector ${\bf q}$ in the AF direction due to spin-wave decay into intra-band particle-hole excitations.}
\end{figure}

For the two-band model with $t_3=t_4=-0.85$, which yields circular electron and hole pockets,\cite{raghu_2008} there is again no finite damping for ${\bf q}$ in the F direction. However, for ${\bf q}$ in the AF direction, we find substantial imaginary part at $\omega=\omega_{\bf q}$, indicating available phase space for spin wave decay into intra-band particle-hole excitations. Fig. 4 shows the spin-wave spectral function in the $(0,\pi)$ state for three different wavevectors. The spin wave linewidth $\Gamma_{\bf q}$ (half-width-at-half-maximum) is seen to increase with spin wave energy, and we obtain $\Gamma_{\bf q}/\omega_{\bf q} \sim 0.25$. Here the spin-wave spectral function was obtained from the RPA-level propagator
as:
\begin{equation}
A_{\bf q}(\omega) = {\rm Im} \frac{1}{1-\lambda_{\bf q}(\omega)}
\end{equation}
from the complex eigenvalue $\lambda_{\bf q}(\omega)$ (corresponding to the spin wave mode) of the matrix $[U][\chi^0({\bf q},\omega)]$ involving the interaction matrix $[U]$ given below Eq. (1) and the particle-hole propagator $[\chi^0({\bf q},\omega)]$ in the two-orbital, two-sublattice basis.\cite{faf2}

\section{Spin-charge coupling mechanism for spin-wave damping}

In view of the vanishing intra-band contribution to spin-wave linewidth obtained above for wavevector in the F direction due to absence of phase space for decay into particle-hole excitations, we will next investigate the contribution to spin-wave damping from the spin-charge coupling mechanism which is important in metallic ferromagnets. Within a systematic expansion of the irreducible particle-hole propagator beyond the RPA, the first order quantum corrections involving self-energy and vertex corrections due to electron-magnon coupling physically correspond to a second-order Raman scattering process involving spin-wave decay into a longer-wavelength intermediate-state spin wave accompanied with internal (majority-spin) charge excitations.\cite{quantum,spch3} Effects of this spin-charge coupling mechanism both on the spin wave energy renormalization and spin wave damping have been studied in detail, and the calculated result $\Gamma_{\bf q}/\omega_{\bf q} \sim 0.1$ is in good agreement with the inelastic neutron scattering measurements of spin wave excitations in ferromagnetic manganites.\cite{zhang_2007,qfklm,qfklm2}

The first-order quantum correction to the irreducible particle-hole propagator (AA term in the sublattice basis) due to the spin-charge coupling mechanism is obtained as:
\begin{equation}
\phi^{(1)}_{AA}  ({\bf q},\omega) = \sum_{\bf k,Q} m_{\bf Q} \Gamma_{\rm sp-chg}^2  
\left ( \frac{1}
{E_{\bf k-q+Q}^{\uparrow +} - E_{\bf k}^{\uparrow -} + \omega_{\bf Q} + \omega  - i \eta } 
\right ) 
\end{equation}
in terms of the SDW-state electronic energies $E_{\bf k}^\sigma$.\cite{faf1} The spin-charge coupling vertex:
\begin{equation}
\Gamma_{\rm sp-chg} = U \left [ \frac{1}{E_{\bf k-q}^{\downarrow \oplus} - E_{\bf k}^{\uparrow \ominus} + \omega - i \eta} 
- \frac{1}{E_{\bf k}^{\downarrow \oplus} - E_{\bf k}^{\uparrow \ominus} + \omega - i \eta} \right ]
\end{equation}
identically vanishes at ${\bf q}=(0,0)$ and $(0,\pi)$ to ensure that the Goldstone mode is explicitly preserved. Here the A-sublattice magnon amplitude:
\begin{equation}
m_{\bf Q} = \frac{1}{2} \left ( \frac{J \alpha_{\bf Q} }{\omega_{\bf Q}} + m \right )
\end{equation}
follows from Eq. (4) in terms of the magnon energy $\omega_{\bf Q}$ in the intermediate state. As the magnon spectrum is dominated by zone boundary modes with $\omega_{\bf Q}\approx J\alpha_{\bf Q}/m$, this amplitude approximately reduces to the magnetization $m$ as in metallic ferromagnets. Also, the majority amplitudes in the SDW state are approximated as 1, and only the corresponding dominant terms --- interband terms (indicated by $\oplus$ and $\ominus$) in the interaction vertex and intra-band particle-hole terms (indicated by + and -) in the charge excitation propagator --- were included in the quantum correction. 

\begin{figure}
\vspace*{-0mm}
\hspace*{0mm}
\psfig{figure=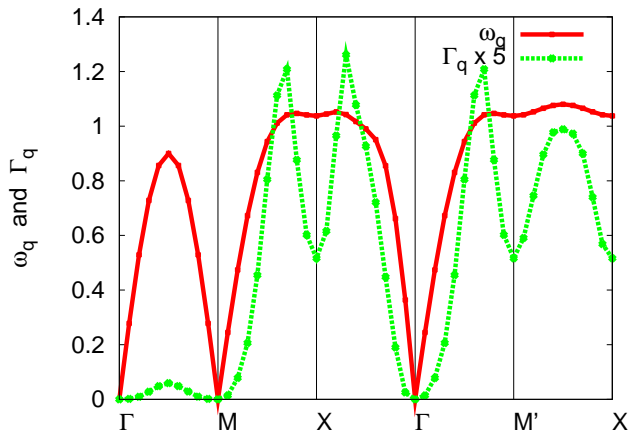,width=85mm,angle=0}
\vspace*{-0mm}
\caption{Momentum dependence of the spin-wave linewidth due to the spin-charge coupling mechanism. Except for the AF direction starting from the zone center, the linewidth is around 10-20\% of the spin wave energy over most of the Brillouin zone.}
\end{figure}

The corresponding spin-wave linewidth due to spin-charge coupling is then obtained from the imaginary part of the self energy $\Sigma_{\rm sw} ({\bf q},\omega) = m U^2 \phi^{(1)} ({\bf q},\omega)$, evaluated at the spin wave energy $\omega = -\omega_{\bf q}$ corresponding to the advanced pole. The $q$-dependence of the spin-wave linewidth $\Gamma_{\bf q}$ is shown in Fig. [5] for the $(0,\pi)$ ordered SDW state of the $t$-$t'$ Hubbard model (with same parameters $\Delta/t=3$, $t'/t=0.3$, $x \approx 40\%$ as earlier). Except in the AF direction from the zone center, over most of the Brillouin zone, the calculated linewidth is about 10-20\% of the spin wave energy, in qualitative agreement with neutron scattering studies of iron pnictides.\cite{zhao_2009} Thus, while spin wave damping in the AF direction is dominated by intra-band particle-hole excitations, it is the spin-charge coupling mechanism which is important for modes in the F direction.  

\section{Conclusions}
In conclusion, spin wave dispersion and damping in iron pnictides can be well understood in terms of physical mechanisms characteristic of metallic magnets, as shown by our comparison with inelastic neutron scattering experiments. The carrier-induced ferromagnetic spins couplings as in metallic ferromagnets are crucial in stabilizing the $(0,\pi)$ and $(0,\pi,\pi)$ ordered metallic SDW states, and analytical expressions for the spin wave dispersion were obtained in two and three dimensions by incorporating these spin couplings in expressions available for the insulating SDW states. On including a small interlayer hopping term, the calculated spin wave dispersion in the $(\pi,0,\pi)$ ordered SDW state was obtained in excellent agreement with experiments. 

With regard to spin wave damping, the intra-band particle-hole excitations and the spin-charge coupling mechanism presented distinctly complementary importance for modes in the AF and F directions in the Brillouin zone. While for modes in the AF direction, the low-energy intra-band particle-hole excitations were found to yield significant linewidth for the two-band model with $\Gamma_{\bf q}/\omega_{\bf q} \sim 1/4$, negligible contribution to linewidth was obtained for modes in the F direction, owing to severe phase-space restriction set by the simultaneous requirements of energy-momentum conservation. On the other hand, the spin-charge coupling mechanism, which is also important in ferromagnetic manganites, was found to yield significant spin wave damping for modes in the F direction, with $\Gamma_{\bf q}/\omega_{\bf q} \sim 10-20\%$, in qualitative agreement with experiments. 

Evidence of Fermi surface folding associated with the SDW state has been observed in recent ARPES studies.\cite{folding_arpes} Electronic quasiparticle dispersion and spectral function renormalization in the SDW states due to electron-magnon interaction and multiple magnon emission-absorption processes should therefore be of interest. 


\end{document}